\documentclass[pra,aps,twocolumn,showpacs,superscriptaddress,tightenlines]{revtex4}

\usepackage{graphicx}
\usepackage{bm}
\usepackage{times}

\begin{document}
\title{Feed-forward and its role in conditional linear optical quantum
dynamics} 

\author{S.~Scheel}\email{s.scheel@imperial.ac.uk}
\affiliation{Quantum Optics and Laser Science, Blackett Laboratory,
Imperial College London, Prince Consort Road, London SW7 2BW, UK} 

\author{W.J.~Munro}
\affiliation{Hewlett-Packard Laboratories, Filton Road, Stoke Gifford,
Bristol BS34 8QZ, UK} 

\author{J.~Eisert}
\affiliation{Quantum Optics and Laser Science, Blackett Laboratory,
Imperial College London, Prince Consort Road, London SW7 2BW, UK} 
\affiliation{Institute for Mathematical Sciences, Imperial College
London, Prince's Gardens, London SW7 2PE, UK}

\author{K.~Nemoto}
\affiliation{National Institute of Informatics, 2-1-2 Hitotsubashi,
Chiyoda-ku, Tokyo 101-8430, Japan} 

\author{P.~Kok}
\affiliation{Hewlett-Packard Laboratories, Filton Road, Stoke Gifford,
Bristol BS34 8QZ, UK} 


\begin{abstract}
Nonlinear optical quantum gates can be created 
probabilistically using only single photon sources, linear optical
elements and photon-number resolving detectors. 
These gates are
heralded but operate with probabilities much less than one. 
There is
currently a large gap between the performance of the 
known circuits
and the established upper bounds on their success 
probabilities. One
possibility for increasing the probability of success of 
such gates
is feed-forward, where one attempts to correct certain 
failure events
that occurred in the gate's operation. In this brief report we 
examine the role of feed-forward in improving the success 
probability. In particular, for the non-linear sign shift gate, 
we find that in a three-mode implementation 
with a single round of feed-forward the optimal average 
probability of success is approximately 
given by $p_{\text{success}}= 0.272$. This value is only slightly
larger than the general optimal success probability
without feed-forward, $p_{\text{success}}= 0.25$.
\end{abstract}

\date{\today}
\maketitle

\section{Introduction}

In recent years we have seen the emergence of photonic states of
light as a possible medium for achieving universal quantum computation
and the medium of choice for quantum communication. Many of the
photon's properties, such as its clean manipulation and negligible
decoherence, makes it ideal to achieve this goal. However, for
scalable quantum
computing
we require photons to interact
with one another. Without this interaction any computation could be
efficiently simulated classically. 
To achieve such interactions it was thought that massive reversible 
nonlinearities were required \cite{Milburn89}. However, materials
giving such large nonlinearities are well beyond our ability to
manufacture. Then, the pioneering work of Knill, Laflamme, and Milburn
(KLM) \cite{KLM} showed that with only single-photon sources,
photon-number resolving detectors, linear optical elements such as
beam splitters, and feed-forward of measurement outcomes,
a near-deterministic controlled-NOT (CNOT) 
gate could be created based on the so-called dual-rail encoding. 
This procedure uses a very significant
fixed overhead of resources, using ``ancilla systems'', 
to achieve some overall failure rate, e.g., one
below the threshold for fault-tolerant quantum computation.
With this architecture for the CNOT gate and single-qubit rotations --
accessible with linear optics -- a  universal set of gates was
possible and a route forward for creating large devices can be
seen
Note that here universal means that an operation
can be implemented that approximates the representation of any unitary
to arbitrary accuracy, based on the dual-rail encoding, where the
logical qubit is encoded in  $|1,0\rangle$ and $|0,1\rangle$. Hence,
it does not mean that any physical gate, as the nonlinear sign shift
gate, can be realized with any probability of success.
Since this original work there has been significant progress
both theoretically
\cite{Pittman01,Knill02,Knill03,Yoran03,Nielsen04,Browne04,Gilchrist05}
and experimentally \cite{Pittman03,OBrien03,Gasparoni04}, with a
number of CNOT gates actually having been demonstrated.

Much of the theoretical effort has focused on determining more
efficient ways to perform the controlled logic. There are two
building blocks of particular importance, the first being the
already mentioned controlled-NOT gate and the second being the
so-called nonlinear sign-shift (NS) gate. This second gate takes a
general two-photon state composed of a superposition of number states
with zero, one, and two photons and flips the sign of the $|2\rangle$
component. So it acts as 
\begin{eqnarray}
c_0 |0 \rangle+c_1|1 \rangle+c_2|2 \rangle\longmapsto
c_0|0 \rangle+c_1|1 \rangle-c_2|2 \rangle,
\end{eqnarray}
where $|n \rangle$ is the $n$-th number state vector of the optical 
field and the coefficients satisfy the usual normalization constraint.
The NS gate is of interest because it is technically more
primitive and fundamental than the CNOT gate, in fact two NS 
gates (in addition to two Hadamard gates) can be used to 
construct a CNOT gate. Using the standard models of linear optical
logic it has been shown in Ref.~\cite{Knill03} that the maximum
probability for achieving the NS gate with postselection is 
bounded from above by $1/2$ 
(and $3/4$ for the CNOT
gate). These upper bounds are known not
to be tight, but they already indicate that
near-deterministic gates are not possible using only the linear
optical resources, toolbox, and strategy. 
Note again that this is no contradiction: using the KLM scheme, 
these non-linear gates can not be
implemented with an arbitrary probability of success, 
but in turn only the representation of any 
unitary in terms of the dual-rail encoding, taken as the
encoding of the logical qubits.

It has 
been shown for small photon numbers in the ancilla system 
\cite{Scheel04a,Scheel05} and later in generality \cite{Eisert04} that
without feed-forward operations (operations that correct situations in 
which the gate has not irrecoverably failed) or the use of 
nonlinear optical resources the maximum probability of success with 
unlimited ancilla is only $1/4$. Ironically, this is just the
value attained in the original proposal in Ref.~\cite{KLM}, 
so this bound is tight. 
This still leaves a lot of space for improvement with
potential appropriate feed-forward steps already on the 
level of NS gates -- with significant
implications on the required overhead in resources in the
scalable scheme. It is manifest that this 
very significant overhead in resources in the full scheme
including feed-forward is indeed 
dictated by the success probability
of the elementary NS gate.

In this article, we will investigate the
possibility of raising 
the success probability of the NS gate using
feed-forward steps already on the level of 
elementary gates, and not only in the full scalable
scheme \footnote{Note that 
it has been recently established
that by expanding the linear optical set to include (weak)
nonlinearities one can construct near-deterministic controlled gates
using homodyne measurements \cite{Nemoto04} or the quantum Zeno effect
\cite{franson04}.}. 
Although we consider only restricted settings
taking the minimal number of auxiliary modes into account, the findings
suggest that even with feed-forward and correction
steps, the success probability cannot be uplifted much at all.

\section{The Nonlinear Sign-Shift gate}

\begin{figure}[ht]
\center{\includegraphics[scale=0.45]{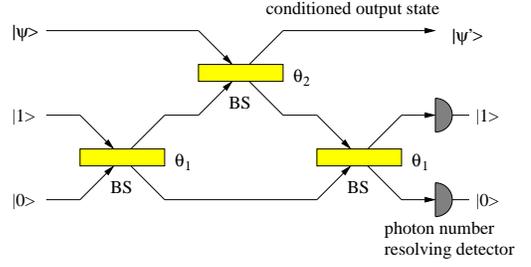}}
 \caption{Schematic diagram of the original KLM nonlinear sign shift
 gate.  The three input states are the unknown two photon signal state
 vector $|\psi\rangle$, plus two ancilla modes, one initially 
 prepared as a single photon $|1\rangle$ and the second as a vacuum
 $|0\rangle$. They interact with each other via the beam splitters
 characterized by $\theta_1,\theta_2\in[0,2\pi)$, respectively. Here,
 $\cos^2 \theta_1=1/(4-2 \sqrt{2})$ and $\cos^2\theta_2=3-2
 \sqrt{2}$. The ancilla modes are then measured using photon-number
 resolving detectors. Upon obtaining the pattern $|1,0\rangle$, the
 signal state vector $|\psi\rangle$ is  transformed 
 to $|\psi'\rangle= c_0 |0\rangle+ c_1 |1\rangle- c_2 |2\rangle$.} 
\label{ns-gate}
\end{figure}

The simplest nonlinear operation/network to be constructed with linear 
optical techniques is the nonlinear sign shift gate originally 
proposed by KLM. An implementation of the gate is 
depicted in Fig.~\ref{ns-gate}, involving the signal mode and two
auxiliary modes. The linear optics network is in this simple
three-mode set-up characterized by two angles
$\theta_1,\theta_2\in[0,2\pi)$,
with $\cos\theta_{1,2}$
denoting the beam splitter transmittivities.
It is straightforward and illustrative to show that for an initial
signal mode input $|n\rangle$ the above gate, conditioned on the
$|1,0\rangle$ detection pattern of the detectors acting on
the two auxiliary modes, with arbitrary angles
$\theta_1$ and $\theta_2$, yields a transformation as \cite{KLM}
\begin{eqnarray}
|0\rangle &\mapsto& \left[ \cos^2 \theta_1 \cos \theta_2 + \sin^2
\theta_1\right]|0\rangle, \nonumber \\  
|1\rangle &\mapsto& - \left[ \cos^2 \theta_1 \cos 2 \theta_2 +
\sin^2 \theta_1 \cos \theta_2\right] |1\rangle, \\  
|2\rangle &\mapsto& \cos \theta_2  \Big[\frac{1}{2}\cos^2
\theta_1 ( 1-3 \cos 2 \theta_2 )
- \sin^2 \theta_1 \cos
\theta_2\Big] |2\rangle. \nonumber  
\end{eqnarray} 
Now with $\theta_1$ and $\theta_2$ chosen such that
$\cos^2 \theta_1=1/(4-2 \sqrt{2})$ and $\cos^2\theta_2=3-2 \sqrt{2}$,
all these three transformations have the same amplitude of
$1/2$ with the $|2\rangle$-component also having a negative
sign. Hence, a general two-photon signal state vector 
$|\psi\rangle$ is transformed according to
\begin{eqnarray}
|\psi\rangle&=& c_0 |0\rangle+ c_1 |1\rangle+ 
c_2 |2\rangle \nonumber \\
&\;&\longmapsto \frac{1}{2} |\psi'\rangle= \frac{1}{2} \left[c_0
|0\rangle+ c_1 |1\rangle-c_2 |2\rangle \right] . 
\end{eqnarray} 
The loss in amplitude reflects the existence of other measurement
outcomes, and so this heralded transformation is effected with a
success probability of 1/4.

\section{Bound on success probabilities}

Let us first examine the limits on the maximum probability of
success. Firstly, for the NS gate Knill \cite{Knill03} has established
a loose upper bound of $1/2$ using a photon-number conservation
argument ($1/4$ is a tight bound without feed-forward 
\cite{Eisert04}).
We will investigate the potential for improvement
by examining the unsuccessful outcomes 
of the gate in a three-mode implementation, that is, 
the situations where we do not measure $|1,0\rangle$. 
These unsuccessful outcomes fit into three categories:
i) Detection of two or more photons in all the ancilla modes.  
   In these cases some information in the signal state is 
   irreversibly destroyed because of photon subtraction.
ii) Detection of  $|0,1\rangle$. In this case the
   photon number in the signal mode has not changed but an incorrect  
   transformation has been applied.
iii) Detection  of $|0,0\rangle$. In this case a
   photon has been added to the signal mode and an incorrect
   transformation has been applied.

The first category contains outcomes where more than one photon is
detected. Consequently, more than zero photons must have been present
in the signal state, and we have therefore obtained information about
the state. Such an error is irrecoverable, as information has leaked out of the
system. The irrecoverable errors
are two-photon patterns -- 
$|1,1\rangle$, $|2,0\rangle$, and $|0,2\rangle$ -- 
and three-photon patterns  -- $|2,1\rangle$, $|1,2\rangle$, $|3,0\rangle$,
and $|0,3\rangle$. 
We need to calculate the probabilities of these
patterns, as their sum gives the total irrecoverable error
probability. This is an indication of the maximum upper probability
bound for the gate to work. The maximum probability of success for the
gate must be less than one minus this irrecoverable error probability. 

To determine the error probabilities we use the techniques 
introduced in Ref.~\cite{Scheel04c} to calculate all the 
necessary matrix elements of any unitary $\hat{U}$ acting
in state space as
\begin{eqnarray}
\label{eq:permanent}
\lefteqn{
\langle m_1,m_2,m_3|\hat{U}|n_1,n_2,n_3\rangle = 
\bigg( \prod\limits_{i,j} m_i!n_j! \bigg)^{-1/2} } \nonumber \\ &&
\times \mbox{per }
\bm{\Lambda}[(1^{m_1},2^{m_2},3^{m_3})|(1^{n_1},2^{n_2},3^{n_3})] \,.
\end{eqnarray}
Here, the multi-index $(1^{m_1},2^{m_2},3^{m_3})$ corresponds to an
index collection in which the index $i$ occurs $m_i$ times. The
symbol ``per'' denotes the {\em permanent} of the unitary
$\bm{\Lambda}$ acting on the bosonic annihilation operators
associated with the unitary $\hat{U}$.

The probability of getting one of the wrong results is
state-dependent, since the corresponding transformation does not
constitute a unitary operation on the signal state. There are then
several different ways of proceeding: One could calculate the
\textit{average} failure rate by averaging the failure probability
over all possible input states. In this case we obtain 
\begin{eqnarray}
\bar{p}_{\text{failure}} = \frac{41}{\sqrt{2}} -\frac{86}{3} \approx
0.325 \,. 
\end{eqnarray}
On the other hand, one can calculate the \textit{maximal} failure rate
by looking at the class of input state for which the failure
probability becomes extremal. This is the case when $c_0=c_1=0$ and
$c_2=1$ in which case we obtain
\begin{eqnarray}
\label{eq:maxfailure}
p_{\text{max,failure}} = 57\sqrt{2}-80 \approx 0.61 \,.
\end{eqnarray}
This failure probability is larger than the suggested maximal failure
rate of $1/2$ in Ref.~\cite{Knill03} which hints at a possible
strictly lower bound on the success probability than $1/2$.  
It is more adequate to consider the maximal failure rate as it
gives truly the worst performance of the gate which is more
appropriate when setting bounds \footnote{In the same spirit, we can
examine the situation in which we have an initial
$|1,1\rangle$-ancilla rather than $|1,0\rangle$. In this situation the
maximum probability of success for the gate is $0.236$ with a
maximal failure probability of $p_{\text{max,failure}} \approx
0.546$ which again exceeds $1/2$.}. Obviously, the
(state-independent) success probability cannot be larger than one
minus the (state-dependent) maximal failure rate. 
So far in our considerations we have looked only at the irrecoverable
errors. There are two other ancilla detection patterns ($|0,1\rangle$
and $|0,0\rangle$) which correspond to incorrect transformations that
do not destroy the information in the signal state.

\section{Correctable error events}

We will briefly look at the cases in which the measurement pattern does
not result in a complete failure, but in a potentially recoverable
error. For simplicity we will assume for the moment that the network
has been tuned to produce the maximal success rate, a condition that
will be relaxed later.
Let us consider first the situation in which no photons are
detected in the ancilla modes, that is, our measurement result
$|0,0\rangle$ occurs from the ancilla $|1,0\rangle$ input. In
this case the (unnormalized) output state vector from the NS 
gate is
\begin{eqnarray}
c_0 2^{-1/4} |1\rangle &+&c_1 2^{1/4}(1-\sqrt{2}) |2\rangle \nonumber \\
&+& c_2 2^{-1/4} (51-36\sqrt{2})^{1/2} |3\rangle \,.
\end{eqnarray}
That is, the information about the input state is still there, but the
state contains too many photons. The smallest term in this equation has an 
amplitude of $2^{-1/4} (51-36\sqrt{2})^{1/2}\approx 0.25$ and so can at
best only increase the success probability for the worst input by
approximately $0.25^2=0.062$ to $0.25 + 0.062 = 0.312$. 
To achieve that, one has to assume the possibility of perfect
recovery, which is unlikely. Hence, let us determine how efficiently we
can recover from this error syndrome. This will be achieved by
applying a second conditional network that subtracts one photon. There 
are several possibilities available to us. The simplest one would be a
single beam splitter with a vacuum input and a single-photon
detection. That, however, does not contain enough parameters to
enable the correction to occur with nonzero probability. 

\begin{figure}[t]
\includegraphics[width=6.5cm]{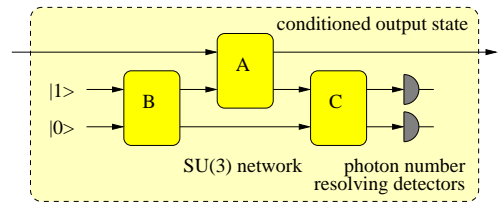}
\caption{A schematic diagram of an SU(3)-network with an initial 
$|10\rangle$-ancilla. The boxes A, B and C represent general SU(2)-networks. }
\label{fig:su3}
\end{figure}

Similarly, an SU(3) network (depicted in Fig.~\ref{fig:su3}) with
$|0,0\rangle$-ancilla and $|1,0\rangle$-detection does not help since
only two of the beam splitters (A and C in Fig.~\ref{fig:su3})
effectively take part in the process. That means, that at least one
photon has to take part in the recovery process and so we must start
with an ancilla of the type $|1,0\rangle$ again. For that we have to
make a choice of the measurement pattern again. If we choose
$|1,1\rangle$, then with the beam splitter angle choices 
$\theta_A=\theta_C=0.489377$, $\theta_B=1.07621$ we end up with a
success probability of only approx. $0.007$ which increases the
overall probability of success of our gate to $0.257$. This
improvement is very small and rather discouraging.

However, there is one more error syndrome we can attempt to
correct. This is the situation in which our measurement result was
$|0,1\rangle$. The number of photons in the probe beam has not changed
but the two-photon input state vector has been transformed into
\begin{eqnarray}
\frac{1-\sqrt{2}}{2}c_0|0\rangle +\frac{5-3\sqrt{2}}{2}c_1|1\rangle
+\frac{15-11\sqrt{2}}{2}c_2|2\rangle \,.
\end{eqnarray}
The smallest amplitude in this state has an amplitude of
$\left|(1-\sqrt{2})/{2}\right|\approx 0.21$ and thus corresponds to a
probability of $0.043$. 
Again, it seems unlikely we can achieve this total correction by linear
optical techniques. Using our general SU(3) network with 
$\theta_A=\theta_C=2.53787$, $\theta_B=2.26111$  we can correct  
this error syndrome with a total probability of $0.015$
(compared to the maximal possible value of $0.043$).
Now adding all the successful error-syndrome-corrected probabilities
together with our $1/4$ initial gate success probability  we get 
\begin{equation}
p_{\text{total success}} \approx 0.272.
\end{equation}
This is a slight improvement but far less than the upper bound
of $1/2$, or even the upper bound $1-0.61=0.39$ from
Eq.~(\ref{eq:maxfailure}). In fact, if we were to take the stance that
we could somehow correct all the recoverable error syndromes, then we
would have had a success probability
$p_{\text{maximal success}} \approx 0.355$. 

\begin{figure}[ht]
\centerline{\includegraphics[width=7cm]{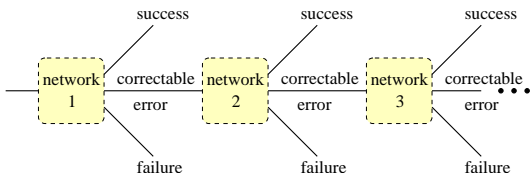}}
\caption{Sequence of conditional networks. Each
application can have one of three possible outcomes: success,
correctable error, or uncorrectable failure.}
\label{fig:sequence} 
\end{figure}

So far we have looked only at a single round of error syndrome
correction. We can of course attempt to correct the recoverable  
errors from the first round (depicted in Fig.~\ref{fig:sequence}).
Unfortunately, a numerical study provides some
evidence that this only changes the total success probability
$p_{\text{total success}}$ by less than one percent and also 
decreases $p_{\text{maximal success}}$ by several percent. 

The result so far is that one or two recovery steps using additional
SU(3) networks do not greatly help to improve the success probability of
the NS gate. We have assumed until now that the first network has been
individually optimized with respect to its probability of success
which seems to be a sensible thing to do experimentally. Let us now
relax this condition. Now we consider in the first instance two SU(3)
networks as in Fig.~\ref{fig:sequence} with three beam splitters each.
Using Eq.~(\ref{eq:permanent}) with the $3\times 3$ unitary
$\bm{\Lambda}$ being a concatenation of the beam splitter matrices
with angles $\theta_A$, $\theta_B$, $\theta_C\in[0,2\pi)$, 
respectively, we
obtain for the matrix elements of
$\hat{U}$
\begin{eqnarray}
\langle 0,1,0|\hat{U}|0,1,0\rangle &=& \mbox{per }\bm{\Lambda}[2|2] =
\Lambda_{22} \,, \\
\langle 1,1,0|\hat{U}|1,1,0\rangle &=& \mbox{per }\bm{\Lambda}[1,2|1,2] =
\Lambda_{11} \Lambda_{22} + \Lambda_{12} \Lambda_{21} \,,\nonumber \\
\langle 2,1,0|\hat{U}|2,1,0\rangle &=&
\mbox{per }\bm{\Lambda}[1,1,2|1,1,2] \nonumber \\ & = & 
\Lambda_{11} (\Lambda_{11} \Lambda_{22} + \Lambda_{12} \Lambda_{21}
+\Lambda_{11} \Lambda_{12}) \,.\nonumber
\end{eqnarray}
Note that in each network two beam splitter angles are
constrained by the requirement of performing a particular gate
operation whereas the third angle determines the probability of
success. Now we maximize the total success probability of the
concatenated networks by optimizing the beam splitter angles in both
networks simultaneously, rather than individually. 
It is worth noting that the success probabilities of obtaining
the outcomes $|1,0\rangle$ and $|0,1\rangle$ after the first network
behave exactly oppositely. That is, as a function of the third beam
splitter angle, the minimum in the probability of finding the
measurement result $|0,1\rangle$ occurs exactly where the probability
of finding $|1,0\rangle$ is maximal. This in turn means that there is a
balance between succeeding with the first network and failure recovery
with a subsequent network. If we denote by $p_x^{(i)}$ the probability
of obtaining the measurement outcome $x$ in the $i$-th network,
numerical optimization yields for the overall probability   
\begin{equation}
p_{\text{total success}} = p^{(1)}_{1,0}+p^{(1)}_{0,1}
p^{(2)}_{1,0} \approx
0.28 \,. 
\end{equation}
Again, this is very low, and suggests 
that there should be a tighter bound
than $1/2$ for single rounds of feed-forward. 
Moreover, the maximal failure probability turns out to be  
$p_{\text{max,failure}} \approx 0.66$, 
so that there is only a chance of at most $1/3$ to generate the
nonlinear sign shift gate even with more subsequent conditional
networks.

\section{Conclusions}

We have shown that it is possible to construct nonlinear sign shift
gates that have a probability of success exceeding $1/4$ with linear
optics, photo-detection and feed-forward. These techniques are
applicable when probabilistic gates fail without divulging information
about the input state. This recycling or syndrome recovery allows a
very modest increase in the overall success probability from $1/4$ to
approx.\ $0.28$. This suggests that the success probability
of the building blocks
can not be significantly reduced  by introducing single feed-forward steps already on the level of the building blocks, 
with the implication of a 
not very large reduction of the overhead in resources in the 
full scalable scheme, in turn making use of feed-forward.
Multiple SU(3) networks will
be required consuming a significant number of signal photons. 
While these resources are constant for a fixed overall 
success probability, they are daunting for an
experimentalist. While our techniques have been applied directly to
the NS gate, a similar analysis can be applied to the other
conditional linear optical gates such as the CNOT gate.

Our results suggests that it is not practical to use feed-forward
operations to correct the error syndromes where no information has
been erased about our quantum states or processes. If we want to 
significantly improve the success probability of these 
elementary gates, we need 
to move outside the linear optical toolbox by, for example,
utilizing other sources of nonlinearity.

\acknowledgments
This work was partially funded by the UK Engineering and Physical
Sciences Research Council (EPSRC), the German
DFG, the EURYI grant, the Japanese JSPS, MPHPT, and 
Asahi-Glass research grants and the European Projects 
RAMBOQ and QAP.


\begin{thebibliography}{20}

\bibitem{Milburn89} G.J.~Milburn, Phys. Rev. Lett. \textbf{62}, 2124 (1989).
\bibitem{KLM} E.~Knill, R.~Laflamme, and G.J.~Milburn, Nature (London)
  \textbf{409}, 46 (2001).
\bibitem{Pittman01} T.B.~Pittman, B.C.~Jacobs, and J.D.~Franson,
  Phys. Rev. A \textbf{64}, 062311 (2001).
\bibitem{Knill02} E.~Knill, Phys. Rev. A \textbf{66}, 052306 (2002).
\bibitem{Knill03} E.~Knill, Phys. Rev. A \textbf{68}, 064303 (2003).
\bibitem{Yoran03} N.~Yoran and B.~Reznik,
  Phys. Rev. Lett. \textbf{91}, 037903 (2003). 
\bibitem{Nielsen04} M.A.~Nielsen, Phys. Rev. Lett. \textbf{93},
  040503 (2004).
\bibitem{Browne04} D.E.~Browne and T.~Rudolph,
  Phys. Rev. Lett. \textbf{95}, 010501 (2005).
\bibitem{Gilchrist05} A.~Gilchrist, A.J.F.~Hayes, and T.C.~Ralph,
  \textit{quant-ph/0505125}. 
\bibitem{Pittman03} T.B.~Pittman, B.C.~Jacobs, M.J.~Fitch, and
  J.D.~Franson, Phys. Rev. A \textbf{68}, 032316 (2003).
\bibitem{OBrien03} J.L.~O'Brien, G.J.~Pryde, A.G.~White, T.C.~Ralph,
  and D.~Branning, Nature (London) \textbf{426}, 264 (2003).
\bibitem{Gasparoni04} S.~Gasparoni, J.W.~Pan, P.~Walther, T.~Rudolph,
  and A.~Zeilinger, Phys. Rev. Lett. \textbf{93}, 020504 (2004).
\bibitem{Scheel04a} S.~Scheel and N.~L\"utkenhaus, New
  J. Phys. \textbf{6}, 51 (2004). 
\bibitem{Scheel05} S.~Scheel and K.M.R.~Audenaert, New
  J. Phys. \textbf{7}, 159 (2005). 
\bibitem{Eisert04} J.~Eisert, Phys. Rev. Lett. \textbf{95}, 040502 (2005).
\bibitem{Nemoto04} K.~Nemoto and W.J.~Munro, Phys. Rev. Lett. \textbf{93},
  250502 (2004); S.D.~Barrett et al.,
  Phys. Rev. A {\bf 71}, 060302(R) (2005).  
\bibitem{franson04} J.D.~Franson, B.C.~Jacobs, and T.B.~Pittman,
  Phys. Rev. A \textbf{70}, 062303 (2004).
\bibitem{Scheel04c} S.~Scheel, \textit{quant-ph/0406127}.
\end{thebibliography}
\end{document}